\documentclass[aps,twocolumn,pra,superscriptaddress,amsmath,showpacs,tightenlines]{revtex4-1}
\usepackage{amssymb}
\usepackage{amsmath}
\usepackage{graphicx}
\usepackage{subfigure}
\usepackage{natbib}
\usepackage{epsfig}
\usepackage{amsfonts}
\usepackage{mathrsfs}
\usepackage{CJK}
\usepackage{xcolor}
\usepackage[toc,page,title,titletoc,header]{appendix}
\begin{document}

\title{Vibration induced transparency: Simulating an optomechanical system  via the cavity QED setup with a movable atom}

\author{Mingzhu Weng}
\affiliation{Center for Quantum Sciences and School of Physics, Northeast Normal University, Changchun 130024, China}
\author{Tian Tian}
\affiliation{School of Science, Changchun University, Changchun 130022, China}
\author{Zhihai Wang}
\email{wangzh761@nenu.edu.cn}
\affiliation{Center for Quantum Sciences and School of Physics, Northeast Normal University, Changchun 130024, China}

\begin{abstract}
We simulate an optomechanical system via a cavity QED scenario with a movable atom and investigate its application in the tiny mass sensing. We find that the steady-state solution of the system exhibits a multiple stability behavior, which is similar to that in the optomechanical system. We explain this phenomenon by the opto-mechanical interaction term in the effective Hamiltonian. Due to the dressed states formed by the effective coupling between the vibration degree of the atom and the optical mode in the cavity, we observe a narrow transparent window in the output field. We utilize this vibration induced transparency phenomenon to perform the tiny mass sensing. We hope our study will broaden the application of the cavity QED system to quantum technologies.
\end{abstract}


\maketitle
\section{introduction}

The coherent mixing of multiple excitation pathways provides the underlying mechanism for many physical phenomena~\cite{MF2005}. A prominent example observed in atomic three-level systems is electromagnetically induced transparency (EIT), in which a weak probe light is transparent when another strong driving field is applied. This phenomenon is caused by the destructive interference among different energy-level transition paths. After this general EIT effect was first observed in cold atomic ensembles~\cite{KJ1991}, its unique applications in nonlinear optics and optical (quantum) information processing have received considerable attention over the past two decades{~\cite{AV2001,YW2004,AR2015}}. The dynamical control of EIT can be used to achieve optical storage~\cite{DF2001,CL2001}, slow-light propagation~\cite{TC2005} and other applications, making it an important part of quantum information and communication proposals, as well as of great practical interest in classical optics and photonics~\cite{MD2001}.

Cavity optomechanical systems exhibit a similar EIT behavior which is named by opto-mechanical induced transparency (OMIT)~\cite{SW2010,AH2011}. This interference effect can be confirmed by the typical $\Lambda$-type three-level system which is formed by the coupling between the mechanical oscillator and optical cavity~\cite{GS2010,JZ2003}. Recently, with the development of nanophotonics and nanofabrication technologies, the application of OMIT in quantum communication has made remarkable progress both experimentally and theoretically~\cite{HX2018}. Furthermore,  OMIT has been found to be manifested in numerous different physical mechanisms, such as multilevel atomic systems~\cite{SL2008,PK2016,PC2014}, nonlinear quantum systems~\cite{KB2013,AK2013}, and hybrid optomechanical systems~\cite{HW2014,YX2014}.

Owing to the rapid development of mechanical resonance, the nanoscale systems are of
great potential application in the mass sensing due to their small size and high sensitivity to the environment~\cite{JL2011,BL2008,KJ2008,CJ2014}. An exciting possibility is the coherent coupling of the mechanical resonators to the condensed matter systems, such as semiconductor quantum dots~\cite{HY2008,JW2010}. In these systems, the OMIT with sensitive transparent windows also provides another way to precisely measure various fundamental physical quantities, for example, the electric charges~\cite{JQ2012}, the magnetic field~\cite{ZX2017} as well as the temperature~\cite{QW2015}. Moreover, the cavity optomechanical system has also been widely used in the field of mass detection{~\cite{DA2003,PD2019,PD2018,KDZhu,QL2017}}

The above achievements show that the optomechanical system has provided an excellent platform for various quantum technologies. Therefore, it is necessary to simulate the optomechanical system with a more simple physical setup. In our previous work, we found that the cavity QED system containing a moving atom can generate an opto-mechanical type interaction~\cite{MW2022} between the vibration degree of freedom of the atom and the cavity mode. This interaction further induces a two-frequency Rabi oscillation. In this paper, we further find that the vibration makes the probe light to the atom
transparent. This vibration induced transparency (VIT), which is the analogy to the EIT and OMIT, is
subsequently used to sense the tiny mass by the distance of two peaks in the output spectrum.

The rest of the paper is organized as follows. In Sec.~\ref{Model}, we describe our theoretical model and  the equations of motion for the system operators. In Sec.~\ref{bistable}, we obtain the steady-state solutions and show the multiple stability characters of the system. In Sec.~\ref{VIT},  we study the VIT in our system and its implementation in the tiny mass sensing. We summarize our results in Sec.~\ref{con}. Some detailed derivations of the steady state solution are given in the Appendix.
\section{Model and Hamiltonian}
\label{Model}
\begin{figure}
\begin{centering}
\includegraphics[width=1\columnwidth]{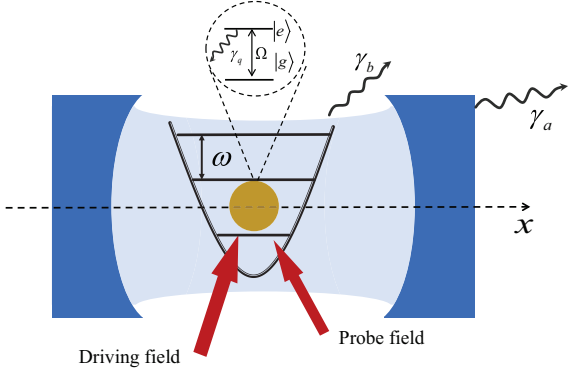}
\par\end{centering}
\caption{Schematic diagram of the model: a single-mode cavity couples a moving two-level atom which is bounded by a harmonic potential.}
\label{device}
\end{figure}

We consider a system illustrated in Fig.~\ref{device}, where a moving but space-constrained two-level atom is coupled to a single-mode cavity. We consider that the spatial motion of the atom is along the $x$ direction which is perpendicular to the cavity wall and is confined by a harmonic potential of oscillator frequency $\omega$. In the simultaneous presence of strong driving and weak probe fields, the Hamiltonian of the total system within rotating wave approximation can be written as $H=H_0+H_d$, where
\begin{eqnarray}
H_0&=&\frac{p^{2}}{2M}+\frac{1}{2}M\omega^{2}x^{2}+\hbar\omega_a a^{\dagger}a+\frac{1}{2}\hbar\Omega\sigma_z\nonumber \\
&&+\hbar g(a^{\dagger}\sigma_- e^{-ikx}+{\rm H.c.}),
\label{h0}
\end{eqnarray}
\begin{equation}
H_d=\hbar\xi(\sigma_{+}e^{-i\omega_{d}t}+{\rm H.c.})+\hbar\epsilon(\sigma_{+}e^{-i\omega_{p}t}+{\rm H.c.}).
\end{equation}
Here, $x$ and $p$ are the atomic position and momentum operators, $a\,({a}^{\dagger})$ is the annihilation (creation) operator of the single-mode cavity field with frequency $\omega_a$, $\sigma_-=(\sigma_+)^{\dagger}=|g\rangle\langle e|$, $\Omega$ is the atomic transition {frequency}, and $M$ is the mass of the atom. $k$ is the photon wave vector in the cavity. $g$ describes the coupling {strength} between the atom and the cavity. $\omega_d$ and $\omega_p$ are the frequencies of the strong driving and  weak probe fields applied to the atom, with $\xi$ and $\epsilon$ ($|\xi|\gg |\epsilon|$) being their Rabi frequencies, respectively.

We introduce the position and momentum representation, $x={X_{\rm zpf}} (b^{\dagger}+b), p=i\hbar(b^{\dagger}-b)/(2{X_{\rm zpf}})$ (${X_{\rm zpf}}=\sqrt{\hbar/2M\omega}$) with $b(b^\dagger)$ denoting the annihilation (creation) operator of the bosonic phonon mode with harmonic oscillator frequency $\omega$. Then, the Hamiltonian in Eq.~(\ref{h0}) can be rewritten as
\begin{eqnarray}
H_0&=&\hbar\omega b^{\dagger}b+\hbar\omega_a a^{\dagger}a+\frac{1}{2}\hbar\Omega\sigma_z\nonumber \\&&+\hbar g[a^{\dagger}\sigma_- e^{-ik{X_{\rm zpf}}(b^{\dagger}+b)}+{\rm H.c.}],
\label{hb}
\end{eqnarray}
where the constant terms have been neglected.

We eliminate the operator on the exponent by applying a unitary transformation  $U=e^{ik {X_{\rm zpf}}(b^{\dagger}+b)a^{\dagger}a}$ to the total Hamiltonian. Then, the effective Hamiltonian $\tilde{H}=UHU^{\dagger}=\tilde{H}_0+\tilde{H}_d$ is obtained as
\begin{eqnarray}
\tilde{H}_0
&=&i\hbar\eta a^{\dagger}a(b-b^{\dagger})+\hbar\chi(a^{\dagger}a)^{2}+\hbar\omega_{a}a^{\dagger}a+\frac{1}{2}\hbar\Omega\sigma_z\nonumber \label{h1}\\
&&+\hbar g(a^{\dagger}\sigma_{-} +a\sigma_{+})+\hbar\omega b^{\dagger}b,\\
\tilde{H}_d&=&\hbar\xi(\sigma_{+}e^{-i\omega_{d}t}+{\rm H.c.})+\hbar\epsilon(\sigma_{+}e^{-i\omega_{p}t}+\rm{H.c.}),
\end{eqnarray}
where
$\chi= k^{2} {X^{2}_{\rm zpf}}\omega$ and $\eta= k{X_{\rm zpf}}\omega$.

The coupling between the vibration degree of freedom of the atom and the cavity field leads to two effects, which are represented by the first and second terms in Eq.~(\ref{h1}). The first term is the effective optomechanical interaction with strength $\eta$. It changes the energy-level transition path of the system compared to the situation with a stationary atom in the traditional cavity QED setup~\cite{MW2022}. The second term of Eq.~(\ref{h1}) represents the Kerr effect with strength $\chi=k^2 {X^{2}_{\rm zpf}}\omega=\hbar k^2/(2M)$.  {Experimentally speaking, the trapping of the atom can be realized by the optical tweezers and the depth of the harmonic trap $\omega$ can be in the order of MHz ~\cite{LTC2012,LH2017}. In contrast to the measurement scheme based on the spatial variation of the field in the cavity QED setup~\cite{TS1993}, the center of mass vibration of the atom is confined by the magneto-optical trap, and there is no spatial vibration of the field in our consideration.}  Taking the parameters in a typical Rydberg atom system, {$ k=10^{7}\,{\rm m^{-1}}, M=10^{-25}\,{\rm kg}, g=1\,{\rm MHz}$~\cite{Anderson2011,Tikman2016,Bounds2018}}, the strength of Kerr interaction is $\chi=0.05g$, which is much weaker than that of the atom-cavity coupling, that is, $\chi\ll g$. Therefore, we are more interested in the novel phenomenon brought about by the effective optomechanical coupling in our work.

Neglecting the Kerr interaction and working in the rotating reference frame with frequency $\omega_d$, the total Hamiltonian approximately becomes
\begin{eqnarray}
H_r&\approx&\hbar\omega b^{\dagger}b+\hbar\Delta_{c}a^{\dagger}a+\frac{1}{2}\hbar\Delta_a\sigma_z\nonumber\\
&&+\hbar g(a^{\dagger}\sigma_{-} +a\sigma_{+})+i\hbar\eta a^{\dagger}a(b-b^{\dagger})\nonumber\\
&&+\hbar\xi(\sigma_{+}+\sigma_{-})+\hbar\epsilon(\sigma_{+}e^{-i\delta t}+\sigma_{-}e^{i\delta t}),
\end{eqnarray}
where $\Delta_{c}=\omega_a-\omega_d\, (\Delta_a=\Omega-\omega_d)$ is the detuning between the cavity field frequency $\omega_a$ (atomic transition frequency $\Omega$) and the driving field frequency $\omega_d$. $\delta=\omega_p-\omega_d$ describes the detuning between the frequency of the weak probe field and the strong driving field.

According to the Heisenberg equations for the motion of the operators and including the dissipation and fluctuation terms, we can obtain the quantum Langevin equations as
\begin{eqnarray}
\frac{d}{dt}a&=&-(i\Delta_{c}+\frac{\gamma_a}{2})a-ig\sigma_{-}+\eta a(b-b^\dagger)\nonumber\\
&&+\sqrt{\gamma_a}A_{\rm{in}},
\label{da}\\
\frac{d}{dt}b&=&-(i\omega+\frac{\gamma_b}{2})b-\eta a^{\dagger}a+\sqrt{\gamma_b}B_{\rm{in}},
\label{db}\\
\frac{d}{dt}\sigma_{-}&=&-(i\Delta_{a}+\frac{\gamma_q}{2})\sigma_{-}+iga\sigma_z\nonumber\\
&&+i\epsilon \sigma_z e^{-i\delta t}+\sqrt{\gamma_q}L_{\rm{in}},
\label{dL}\\
\frac{d}{dt}\sigma_z&=&-\gamma_{q}(\sigma_z +1)+2ig(a^{\dagger}\sigma_{-} -a\sigma_{+})+2i\xi(\sigma_{-}-\sigma_{+})\nonumber\\
&&+2i\epsilon (\sigma_{-}e^{i\delta t}-\sigma_{+}e^{-i\delta t})+\sqrt{\gamma_q}Z_{\rm{in}}.
\label{dZ}
\end{eqnarray}
Here $\gamma_{a},\gamma_{b}$ and $\gamma_{q}$ are the decay rates of the cavity field, atomic vibration mode and the two-level atom, respectively. In addition, $A_{\rm{in}}, B_{\rm{in}}, L_{\rm{in}}$ and $Z_{\rm{in}}$ are the noise operators associated with the photon mode $a$, vibration mode $b$ and atomic operators $\sigma_{-}$, $\sigma_z$, respectively, with zero average values $\langle A_{\rm{in}}\rangle=\langle B_{\rm{in}}\rangle=\langle L_{\rm{in}}\rangle= \langle Z_{\rm{in}}\rangle=0$.

\section{Average values in the steady state}
\label{bistable}

In this section, we will discuss the behavior of the average values in the steady state. According to Eqs.{~(\ref{da})-~(\ref{dZ})} and under the mean-field approximation, i.e., $\langle a^{\dagger} a\rangle=\langle a^{\dagger}\rangle\langle a \rangle$, we have the average value equations for the operators $a,b,\sigma_{-}$ and $\sigma_{z}$ as
\begin{eqnarray}
\langle \dot{a}\rangle&=&-(i\Delta_{c}+\frac{\gamma_a}{2})\langle a\rangle-ig\langle\sigma_{-}\rangle+\eta \langle a\rangle(\langle b\rangle-\langle b^\dagger \rangle),
\label{meanda}\\
\langle \dot{b}\rangle&=&-(i\omega+\frac{\gamma_b}{2})\langle b\rangle-\eta \langle a^{\dagger}\rangle \langle a\rangle,
\label{meandb}\\
\langle \dot{\sigma_{-}}\rangle&=&-(i\Delta_{a}+\frac{\gamma_q}{2})\langle\sigma_{-}\rangle+ig\langle a\rangle\langle\sigma_z\rangle+i\epsilon \langle\sigma_z \rangle e^{-i\delta t},
\label{meandL}\\
\langle \dot{\sigma_{z}}\rangle&=&-\gamma_{q}(\langle\sigma_z\rangle +1)+2ig(\langle a^{\dagger}\rangle\langle\sigma_{-}\rangle -\langle a\rangle\langle\sigma_{+}\rangle)\nonumber\\
&&+2i\xi(\langle\sigma_{-}\rangle-\langle\sigma_{+}\rangle)+2i\epsilon (\langle\sigma_{-}\rangle e^{i\delta t}-\langle\sigma_{+}\rangle e^{-i\delta t}).\nonumber\\
\label{meandZ}
\end{eqnarray}
It is impossible to obtain an exact analytical solution to this set of nonlinear equations because the steady-state response contains infinitely many components with different frequencies. Retaining to the first order sideband, the average values of the system variables can be written approximately as~\cite{RW}
\begin{eqnarray}
\langle a\rangle&=&A_{0}+A_{+}e^{i\delta t}+A_{-}e^{-i\delta t},
\label{solutiona}\\
\langle b\rangle&=&B_{0}+B_{+}e^{i\delta t}+B_{-}e^{-i\delta t},
\label{solutionb}\\
\langle \sigma_{-}\rangle&=&L_{0}+L_{+}e^{i\delta t}+L_{-}e^{-i\delta t},
\label{solutionL}\\
\langle \sigma_{z}\rangle&=&Z_{0}+Z_{+}e^{i\delta t}+Z_{-}e^{-i\delta t}.
\label{solutionZ}
\end{eqnarray}
Here, the zero-order quantities $A_{0},B_{0},L_{0}$, and $Z_{0}$ correspond to the response to the driving field and the first-order quantities $A_{\pm},B_{\pm},L_{\pm}$, and $Z_{\pm}$ represent the response to the probe field. Combining Eqs.~(\ref{solutiona})-(\ref{solutionZ}) with Eqs.~(\ref{meanda})-(\ref{meandZ}), and ignoring the second-order small terms, the above parameters can be obtained by comparing the prefactors in terms of the exponentials $e^{\pm i\delta t}$.

To analyze the stability of the system, we obtain the steady-state value of $|A_{0}|$ as ({see Appendix for the detailed derivations })
\begin{equation}
|A_{0}|=\frac{2\xi g|Z_{0}|(\gamma^{2}_{b}+4\omega^{2})}{\sqrt{\epsilon^{2}_{1}+\epsilon^{2}_{2}}},
\label{meanvalueA}
\end{equation}
Here, the parameters $\epsilon_{1}$ and $\epsilon_{2}$ are given by
\begin{eqnarray}
\epsilon_{1}&=&(\gamma^{2}_{b}+4\omega^{2})(-2\Delta_{a}\Delta_{c}+\frac{\gamma_{q}\gamma_{b}}{2}-2g^{2}Z_{0})\nonumber\\
&&+16\eta^{2}\omega\Delta_{a}|A_{0}|^{2},\\
\epsilon_{2}&=&(\gamma^{2}_{b}+4\omega^{2})(\Delta_{c}\gamma_{q}+\frac{\gamma_{b}\Delta_{a}}{2})-8\eta^{2}\omega\gamma_{q}|A_{0}|^{2}.
\end{eqnarray}

\begin{figure}
\begin{centering}
\includegraphics[width=1\columnwidth]{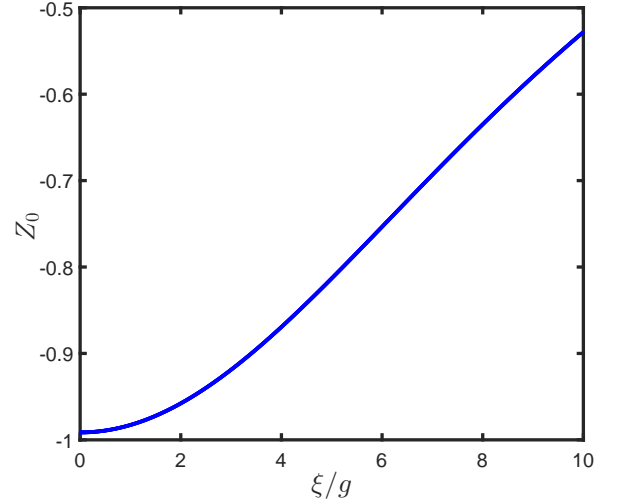}
\par\end{centering}
\caption{$Z_{0}$  as a function of the driving field strength $\xi$. The parameters are set as {$g=1\,{\rm MHz},\omega=0.2\,{\rm MHz},\gamma_{q}=50\,{\rm kHz},\Delta_{a}=10\,{\rm MHz}$ and $\Delta_{c}=0.5\,{\rm MHz}$.}}
\label{Z}
\end{figure}

Generally speaking, under the condition that the frequency of the cavity and the atom is largely {detuned}, that is, $|\Delta_{a}-\Delta_{c}|\gg g$, it is reasonable to consider that the atom is nearly frozen  in the ground state. Combining Eq.~(\ref{meanvalueA}) with Eq.~(\ref{meandZ}) and Eqs.~(\ref{solutiona})-(\ref{solutionZ}), we can obtain the solution
\begin{equation}
Z_{0}=-\frac{4\Delta_{a}^{2}+\gamma_{q}^{2}}{4\Delta^{2}_{a}
+\gamma_{q}^{2}+8g^{2}|A_{0}|^{2}+8\xi^{2}}.
\label{Z0}
\end{equation}

In Fig.~\ref{Z}, we plot the influence of the driving strength on the two-level system according to Eq.~(\ref{Z0}). It shows that $Z_{0}$ is very close to $-1$ when the driving strength $\xi$ locates in the range of $0-2g$.  The result is consistent with our assumption that the atom is nearly frozen in the ground state in the weak driving situation.

\begin{figure}
\begin{centering}
\includegraphics[width=1\columnwidth]{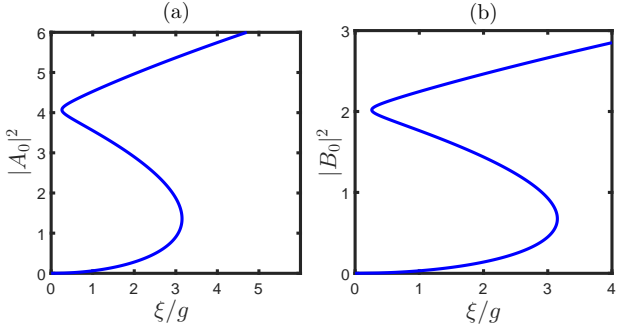}
\par\end{centering}
\caption{The steady-state values of (a) $|A_{0}|^{2}$  and (b) $|B_{0}|^{2}$ as functions of the driving strength $\xi$.  The parameters are set as $Z_{0}=-0.99, \gamma_{a}=60\,{\rm kHz}, \gamma_{b}=50\,{\rm kHz}, \gamma_{q}=50\,{\rm kHz}$.}
\label{bistablefig}
\end{figure}

Furthermore, in Fig.~\ref{bistablefig}(a), we plot the steady-state photon number $|A_{0}|^{2}$ as a function of the {driving} strength $\xi$. This is very similar to the bistable phenomenon which is caused by the nonlinear effects in other cavity hybrid systems~\cite{SM,HW,RG}. From Eq.~(\ref{meanvalueA}), we can also observe a consistent conclusion that $|A_{0}|^{2}$ has three solutions when $Z_{0}$ takes a constant.

{If we exclude the effect of the effective opto-mechanical coupling by setting $\eta=0$, Eq.~(\ref{meanvalueA}) becomes}
\begin{equation}
{|A_{0}|=\frac{2\xi g |Z_{0}|}{\sqrt{(-2\Delta_{a}\Delta_{c}+\frac{\gamma_{q}\gamma_{b}}{2}-2g^{2}Z_{0})^2+(\Delta_{c}\gamma_{q}+\frac{\gamma_{b}\Delta_{a}}{2})^2}}}.
\end{equation}

{Therefore, $|A_{0}|^2$ has only one unique solution, which means that the system does not exhibit multiple stability behavior.  This indicates that the multiple stability is induced only by the effective optomechanical interaction term. In our work, the linearization under weak driving conditions ruled out the possibility that the two-level atom act as the nonlinear element to cause the bistable behavior of the system~\cite{nonlinearity}. The situation for strong driving where the linearization fails is beyond our consideration in this paper.}

{We also obtain the steady-state value for the vibration degree of freedom of the atom as }
\begin{equation}
{B_{0}=\frac{-2\eta|A_{0}|^{2}}{\gamma_{b}+2i\omega}},
\end{equation}
{which linearly depends on the steady-state value $|A_{0}|^2$ of the cavity field. In Fig.~\ref{bistablefig}(b), the steady-state phonon number $|B_{0}|^{2}$ is plotted as a function of the Rabi frequency $\xi$, which also shows the  bistabilities.}

\section{Vibration induced transparency and mass sensing}
\label{VIT}
\subsection{Vibration Induced Transparency}

We now study the response of the system to the weak probe field, which can be detected by the output field. Using the input-output theory~\cite{DF}
\begin{equation}
\langle L_{\rm{out}}\rangle=\sqrt{\gamma_{q}}\langle \sigma_{-}\rangle-\frac{\xi}{\sqrt{\gamma_{q}}}-\frac{\epsilon}{\sqrt{\gamma_{q}}}e^{-i\delta t},
\end{equation}
we assume
\begin{equation}
\langle L_{\rm{out}}\rangle=L_{\rm{out}0}+L_{\rm{out}-}\epsilon e^{-i\delta t}+L_{\rm{out}+}\epsilon^{*}e^{i\delta t}.
\label{Lout}
\end{equation}
Substituting Eq.~(\ref{solutionL}) into the above equation, we can get
\begin{eqnarray}
L_{\rm{out}0}&=&\sqrt{\gamma_{q}}L_{0}-\frac{\xi}{\sqrt{\gamma_{q}}},\\
L_{\rm{out}-}&=&\frac{\sqrt{\gamma_{q}}}{\epsilon}L_{-}-\frac{1}{\sqrt{\gamma_{q}}},\\
L_{\rm{out}+}&=&\frac{\sqrt{\gamma_{q}}}{\epsilon^{*}}L_{+}.
\end{eqnarray}
Then, the coefficients $L_{0}$ and $L_{\pm}$ are obtained as (see the appendix for the detailed derivations)
\begin{eqnarray}
L_{0}&=&\frac{2gA_{0}Z_{0}+2\xi Z_{0}}{2\Delta_{a}-i\gamma_{q}},\\
L_{+}&=&\frac{L_{2}L_{6}^{*}+L_{3}(1-L_{4}^{*})}{(1-L_{1})(1-L_{4}^{*})-L_{2}L_{5}^{*}},\\
L_{-}&=&\frac{L_{3}L_{5}^{*}+L_{6}(1-L_{1}^{*})}{(1-L_{1}^{*})(1-L_{4})-L_{2}^{*}L_{5}}.
\end{eqnarray}
From Eq.~(\ref{Lout}), {$L_{\rm{out}-}$ corresponds to the response of the system to the probe field.} Therefore, we will study the effect of $L_{-}$ to understand the behavior of the output field. We redefine the output field at frequency $\omega_{p}$ of the probe field as $\epsilon_{\rm{out}}=\gamma_{q}L_{-}/\epsilon$ with the real and imaginary parts
\begin{eqnarray}
\mu_{p}&=&\frac{\gamma_{q}(L_{-}+L_{-}^{*})}{2\epsilon},\\
\nu_{p}&=&\frac{\gamma_{q}(L_{-}-L_{-}^{*})}{2i\epsilon}.
\end{eqnarray}

\begin{figure}
\begin{centering}
\includegraphics[width=1\columnwidth]{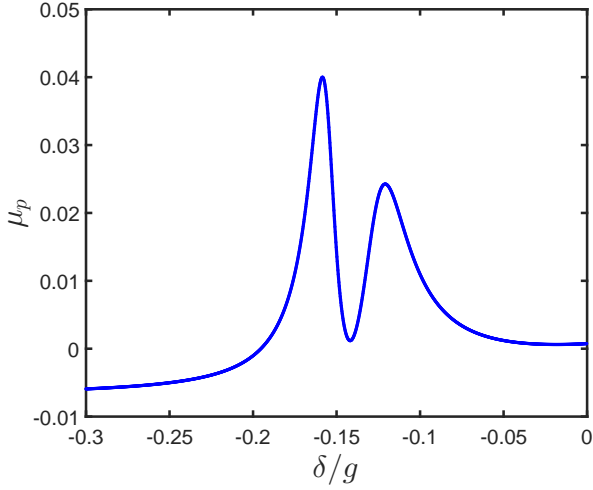}
\par\end{centering}
\caption{The real part $\mu_{p}$ of the amplitude $\epsilon_{\rm{out}}$ of the output field varies with the detuning $\delta$. The parameters are set as {$M=10^{-25}{\rm kg}, \Delta_{a}=10 {\rm MHz}, \Delta_{c}=0.09 {\rm MHz},\gamma_{q}=0.3{\rm MHz}, \gamma_{a}=8{\rm kHz},\gamma_{b}=1\,{\rm kHz}$ and $\xi=9{\rm MHz}$. }}
\label{eit}
\end{figure}

It is clear that $\mu_{p}$ represents the absorption spectrum of the output field while $\nu_{p}$ describes its dispersion. In Fig.~\ref{eit}, we plot the absorption spectrum of the probe field within the experimental available parameters. We can observe a similar transparent window as that in the optomechanical system. This window is induced by the vibration of the atom and we name it VIT, in analogy with OMIT in a real optomechanical system~\cite{SW2010,HW2014}.

To gain more physical insight into VIT, it is instructive to transfer to
the sideband representation as shown in Fig.~\ref{energy}. First of all, for the convenience of description, we define the state $|m,n,\sigma\rangle:=|m\rangle_c\otimes|n\rangle_v\otimes|\sigma\rangle_a$ ($|\sigma\rangle=|e\rangle,|g\rangle$) to represent that the cavity mode (vibrate mode) is in the bosonic Fock state with $m (n)$ excitations while the atom is in the state $|\sigma\rangle$.

The eigen wave function of $\tilde{H}_0$ without effective optomechanical coupling can be obtained as
\begin{eqnarray}
|\psi_+^{(m,n)}\rangle&=&\cos\frac{\theta}{2}|m,n,e\rangle+\sin\frac{\theta}{2}|m+1,n,g\rangle,\\
|\psi_-^{(m,n)}\rangle&=&-\sin\frac{\theta}{2}|m,n,e\rangle+\cos\frac{\theta}{2}|m+1,n,g\rangle,
\end{eqnarray}
{where $\tan\theta=2\sqrt{m+1}g/(\Omega-\omega_{a})$ and the optomechanical interaction will further couple them. For example, the states $|\psi_{\pm}^{(0,0)}\rangle$ will couple to states $|\psi_{\pm}^{(0,1)}\rangle$ simultaneously, that is, it forms four transition channels as shown in Fig.~\ref{energy}. However, in the parameter regime we consider, the coupling between $|\psi_{+}^{(0,0)}\rangle$ and $|\psi_{-}^{(0,1)}\rangle$  denoted by the red line in Fig.~\ref{energy}(a) will play the most important role due to its smallest energy spacing. Therefore, for a brief discussion, we only consider the transitions in this channel and the transition strength $\kappa$ is obtained as
\begin{eqnarray}
\hbar \kappa\approx\langle\psi_{+}^{(0,0)}|\hbar\eta a^{\dagger}a(b-b^{\dagger})|\psi_{-}^{(0,1)}\rangle
=\frac{1}{2}\hbar\eta.
\end{eqnarray}
It then forms another two dressed states $|\Psi_\pm\rangle$,  as shown in Fig.~\ref{energy}(b), which are the superposition of $|\psi_+^{(0,0)}\rangle$ and $|\psi_-^{(0,1)}\rangle$. Therefore, we here achieve a {$V$}-type three-level system which is composed of the states $|\psi_-^{(0,0)}\rangle, |\Psi_-\rangle$ and $|\Psi_+\rangle$. The corresponding transition frequency $\omega_{\pm}$ can be obtained as
\begin{eqnarray}
\hbar\omega_{\pm}&=&E_{\pm}-E_-^{(0,0)}\nonumber\\
&=&\hbar g+\frac{1}{2}\hbar\omega\pm\hbar\sqrt{(g-\frac{1}{2}\omega)^{2}+{\kappa^{2}}}.
\end{eqnarray}
with
\begin{eqnarray}
E_-^{(0,0)}&=&\hbar(\omega_a-g),\nonumber\\
E_{\pm}&=&\hbar\omega_a+\frac{1}{2}\hbar\omega\pm\hbar\sqrt{(g-\frac{1}{2}\omega)^{2}+{\kappa^{2}}}.
\end{eqnarray}

\begin{figure}
\begin{centering}
\includegraphics[width=1\columnwidth]{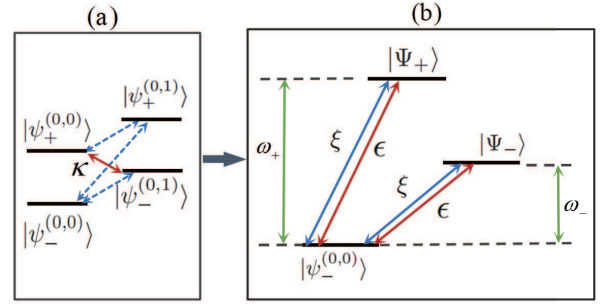}
\par\end{centering}
\caption{(a) The original simplified energy-level diagram. (b) Schematic diagram for the energy-level transition between different sidebands with the driving and probing fields.}
\label{energy}
\end{figure}

{The driving and probing to the atom induces the transitions from $|\psi_{-}^{(0,0)}\rangle$ to $|\Psi_{+}\rangle$ and $|\Psi_{-}\rangle$ and there exist two interference paths in our system. One path is $|\psi_{-}^{(0,0)}\rangle\rightarrow|\Psi_{+}\rangle\rightarrow|\psi_{-}^{(0,0)}\rangle\rightarrow|\Psi_{-}\rangle$ while the other is $|\psi_{-}^{(0,0)}\rangle\rightarrow|\Psi_{-}\rangle$. The transparent window  comes from the destructive interference of these
two paths.} {For the parameters we are considering, we get $|\omega_{-}|\approx 0.18g$, which is consistent with the transparent window in the Fig.~\ref{eit}.} In other words, the transparent window arises from the destructive interference of transition pathways when the detuning satisfies $\delta=\omega_{-}$. {The VIT here is different from those in the standard optomechanical system and the traditional cavity QED system. First, the frequency $\omega_{-}$ is not the natural physical parameters. This is because $|\Psi_{+}\rangle$ and $|\Psi_{-}\rangle$  are the hybrid energy levels resulting from the opto-mechanical interaction, instead of the original bare energy levels of the free system. Moreover, the system under our consideration is subject to a $V$-type three-level structure instead of $\Lambda$-type in OMIT.} Now, we will consider the further practical applications using this VIT phenomenon.

\subsection{Mass Sensing}

\begin{figure}
\begin{centering}
\includegraphics[width=1\columnwidth]{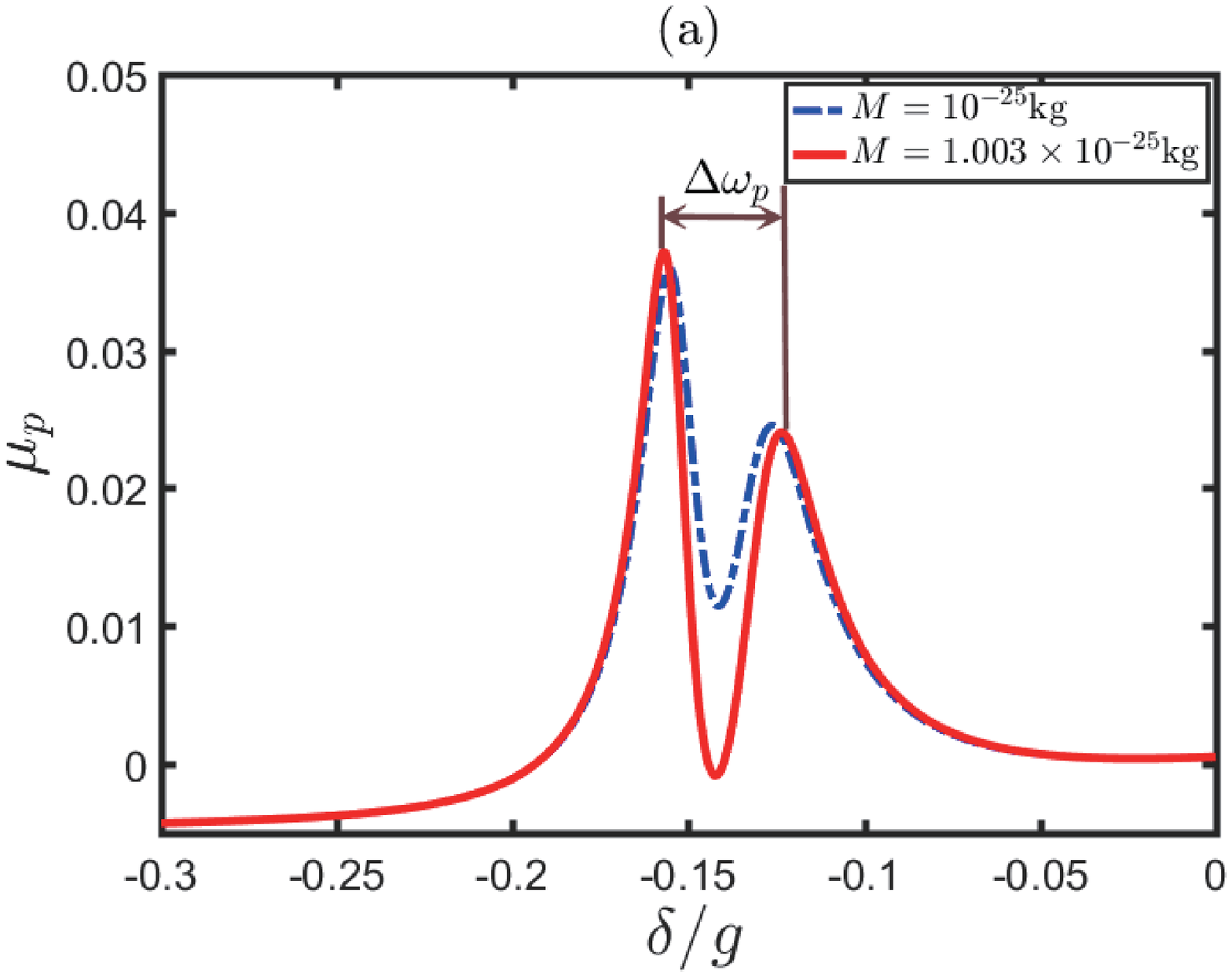}
\includegraphics[width=1\columnwidth]{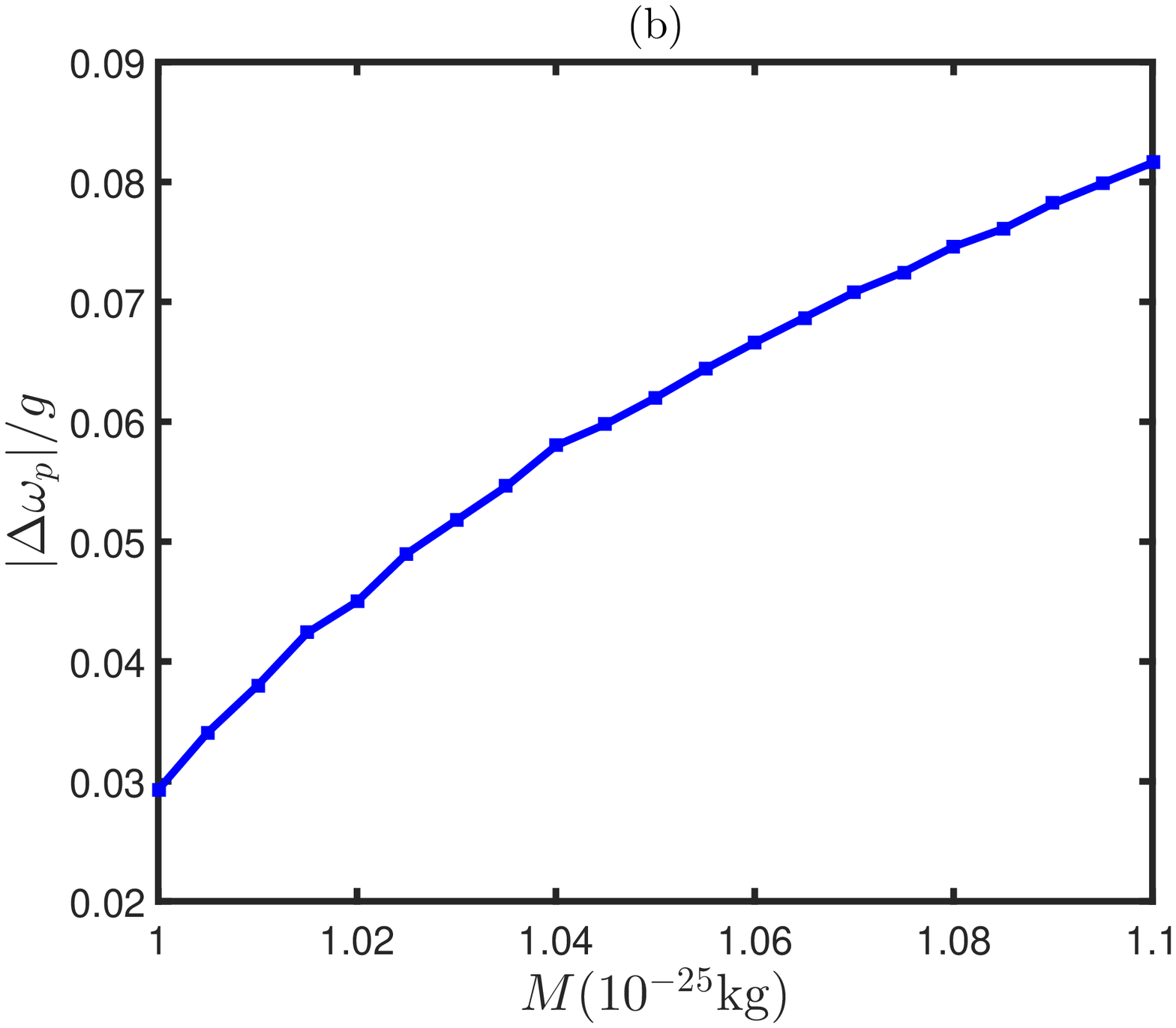}
\par\end{centering}
\caption{(a) The absorption spectrum of the probe field as a function of the detuning between the probe and driving fields for different atomic mass. (b) The width $\Delta\omega_{p}$ of the transparent window varies with the atomic mass $M$. The parameters are set to be same with those in Fig.~\ref{eit}.}
\label{sensing}
\end{figure}

As mentioned above, the narrow VIT window can be observed in the absorption spectrum thanks to the effective optomechanical interaction term. We also note that the effective optomechanical interaction strength $\eta=k{X_{\rm zpf}} \omega\,({X_{\rm zpf}}=\sqrt{\hbar/2M\omega})$ is related to the atomic mass. Therefore, the VIT provides us with a theoretically feasible scheme for the tiny mass sensing. The probe absorption spectrum as a function of the detuning between a probe field and driving field with different atomic mass is shown in Fig.~\ref{sensing}(a). It demonstrates that the splitting $\Delta\omega_p$ between the two peaks in the output spectrum changes with the atomic mass. Therefore, we can measure this splitting variation to sense the tiny mass.  In Fig.~\ref{sensing}(b), we plot the splitting $\Delta \omega_p$ as a function of the deposited mass. It is clear that the splitting is linearly dependent on the atomic mass, indicating that our scheme can be used to detect mass change at the atomic scale.

{At last, we would like to point out the difference between our mass sensing scheme and that in the optomechanical system. For the scheme of the mass sensing scheme in an optomechanical system platform, the change of mass is manifested by the frequency of the phonon. Therefore, the change in the position of the absorption peak reflects the increase or decrease of the atomic mass~\cite{KDZhu}. In our model, the changes of the atomic
mass do not cause the changes of the vibrational frequency, but the strength optomechanical interaction.}

\section{Conclution}
\label{con}
In this paper, we consider a cavity QED system, which is consist of a single-mode field and a driven movable two-level atom. By introducing a harmonic potential to limit the spatial movement of the atom, it is demonstrated that the vibration of the atom induces an efficient optomechanical coupling. We find that the system exhibits a multiple stability behavior due to the quantum transitions which are induced by the effective optomechanical interaction. This prompts us to further study the absorption spectrum of the system which is characterized by the VIT phenomenon.  Furthermore, we theoretically propose a mass sensing scheme by measuring the splitting between the two peaks in the VIT spectrum.  We expect our results will lead to novel developments in the field of cavity QED systems and the atomic scale sensing.
\begin{acknowledgments}
This work is supported by National Key R\&D Program of China (Grant
No. 2021YFE0193500) and National Natural Science Foundation
of China (Grants No. 12105026 and No. 11875011).
\end{acknowledgments}

\appendix
\addcontentsline{toc}{section}{Appendices}\markboth{APPENDICES}{}
\begin{subappendices}
\section{Calculation of $L_{-}$}
\label{Appendix}

In Sec.~\ref{bistable}, we give the approximate solutions to Eqs.~(\ref{solutiona})-(\ref{solutionZ}) for the average values. We substitute the expressions $\langle a\rangle$ and $\langle b\rangle$, which are given by Eqs.~(\ref{solutiona})-(\ref{solutionb}), into Eq.~(\ref{meandb}), then $B_{0}$ and $B_{\pm}$ can be expressed in terms of $ A_{0}, A_{\pm}$ as
\begin{eqnarray}
B_{0}&=&\frac{-2\eta|A_{0}|^{2}}{\gamma_{b}+2i\omega},\\
B_{+}&=&B_{1}(A_{0}^{*}A_{+}+A_{0}A_{-}^{*}),\\
B_{-}&=&B_{2}(A_{0}^{*}A_{-}+A_{0}A_{+}^{*}).
\end{eqnarray}
with
\begin{eqnarray}
B_{1}&=&\frac{-2\eta}{\gamma_{b}+2i(\omega+\delta)},\\
B_{2}&=&\frac{-2\eta}{\gamma_{b}+2i(\omega-\delta)}.
\end{eqnarray}

Similarly, we obtain the steady state average values of $A_{0}$ and $A_{+},A_{-}$, which can be expressed by $L_{+},L_{-}$ with 
\begin{eqnarray}
A_{0}&=&\frac{gL_{0}}{-\Delta_{c}+i\frac{\gamma_{a}}{2}-i\eta(B_{0}-B^{*}_{0})},\\
A_{+}&=&A_{1}L_{+}+A_{2}L_{-}^{*},\label{A+}\\
A_{-}&=&A_{3}L_{-}+A_{4}L_{+}^{*}.\label{A-}
\end{eqnarray}
In the above equations, the parameters are given by
\begin{eqnarray}
A_{1}&=&\frac{gA_{6}^{*}}{A_{5}A_{6}^{*}-\eta^{2}(B_{1}-B_{2}^{*})(B_{2}^{*}-B_{1})|A_{0}|^{2}},\\
A_{2}&=&\frac{i\eta g(B_{1}-B_{2}^{*})}{A_{5}A_{6}^{*}-\eta^{2}(B_{1}-B_{2}^{*})(B_{2}^{*}-B_{1})|A_{0}|^{2}},\\
A_{3}&=&\frac{gA_{5}^{*}}{A_{5}^{*}A_{6}-\eta^{2}(B_{1}-B_{2}^{*})(B_{2}^{*}-B_{1})|A_{0}|^{2}},\\
A_{4}&=&\frac{i\eta g(B_{2}-B_{1}^{*})}{A_{5}^{*}A_{6}-\eta^{2}(B_{1}-B_{2}^{*})(B_{2}^{*}-B_{1})|A_{0}|^{2}},\\
A_{5}&=&-(\delta+\Delta_{c})+i\frac{\gamma_{a}}{2}\nonumber\\
&&-i\eta[(B_{0}-B_{0}^{*})+(B_{1}-B_{2}^{*})A_{0}],\\
A_{6}&=&\delta-\Delta_{c}+i\frac{\gamma_{a}}{2}\nonumber\\
&&-i\eta[(B_{0}-B_{0}^{*})+(B_{2}-B_{1}^{*})A_{0}].
\end{eqnarray}

By substituting the expressions of $\langle a\rangle$, $\langle \sigma_{-}\rangle$ and $\langle \sigma_{z}\rangle$ into the equation of motion of the average value of the operator $\sigma_{z}$, we obtain the expressions of $Z_{+}$ and $Z_{-}$ with Eqs.~(\ref{A+})-(\ref{A-}) as
\begin{eqnarray}
Z_{+}&=&Z_{1}L_{+}+Z_{2}L_{-}^{*}+\lambda_{1}\epsilon L_{0},\\
Z_{-}&=&Z_{3}L_{-}+Z_{4}L_{+}^{*}+\lambda_{1}^{*}\epsilon L_{0}^{*}.
\end{eqnarray}
Here the coefficients $Z_{i}(i=1,2,3,4)$ are found as
\begin{eqnarray}
Z_{1}&=&\lambda_{1}[(gA_{0}^{*}+\xi)+gA_{4}^{*}L_{0}-gA_{1}L_{0}^{*}],\\
Z_{2}&=&\lambda_{1}[-(gA_{0}+\xi)+gA_{3}^{*}L_{0}-gA_{2}L_{0}^{*}],\\
Z_{3}&=&\lambda_{1}^{*}[-(gA_{0}^{*}+\xi)+gA_{3}L_{0}^{*}-gA_{2}^{*}L_{0}],\\
Z_{4}&=&\lambda_{1}^{*}[(gA_{0}^{*}+\xi)+gA_{4}L_{0}^{*}-gA_{1}^{*}L_{0}].
\end{eqnarray}
with $\lambda_{1}=2/\delta-i\gamma_{q}$.
Using the similar steps as above, we can obtain the formulas for $L_{+}$ and $L_{-}$ for the average value of the operator $\sigma_{-}$. Up to the first order of $\epsilon$, we will obtain
\begin{eqnarray}
L_{+}&=&\frac{L_{2}L_{6}^{*}+L_{3}(1-L_{4}^{*})}{(1-L_{1})(1-L_{4}^{*})-L_{2}L_{5}^{*}},\\
L_{-}&=&\frac{L_{3}L_{5}^{*}+L_{6}(1-L_{1}^{*})}{(1-L_{1}^{*})(1-L_{4})-L_{2}^{*}L_{5}},
\end{eqnarray}
which are Eq. (29) and Eq. (30) in the main text. The parameters $L_{i}(i=1,2,...,6)$ in the above equations are
\begin{eqnarray}
L_{1}&=&\lambda_{2}[Z_{1}(gA_{0}+\xi)+gZ_{0}A_{1}],\\
L_{2}&=&\lambda_{2}[Z_{2}(gA_{0}+\xi)+gZ_{0}A_{2}],\\
L_{3}&=&\lambda_{1}\lambda{2}\epsilon L_{0}(gA_{0}+\xi),\\
L_{4}&=&\lambda_{3}[Z_{3}(gA_{0}+\xi)+gZ_{0}A_{3}],\\
L_{5}&=&\lambda_{3}[Z_{4}(gA_{0}+\xi)+gZ_{0}A_{4}],\\
L_{6}&=&\lambda_{3}[(\lambda_{1}^{*}\epsilon L_{0}^{*})(gA_{0}+\xi)+\epsilon Z_{0}],
\end{eqnarray}
with
\begin{eqnarray}
\lambda_{2}&=&\frac{2}{2(\delta+\Delta_{a})-i\gamma_{q}},\\
\lambda_{3}&=&\frac{2}{2(\Delta_{a}-\delta)-i\gamma_{q}}.
\end{eqnarray}

\end{subappendices}

\end{document}